\documentclass[12pt,preprint]{aastex}

\shorttitle{Grain size evolution in AGB stars}
\shortauthors{Speck et al.}

\begin{document}


\title{The effect of stellar evolution on SiC dust grain sizes} 


\author{Angela K. Speck}
\author{Grant D. Thompson}
\affil{Department of Physics \& Astronomy, University of Missouri, Columbia, 
MO 65211}

\and

\author{Anne M. Hofmeister}
\affil{Department of Earth \& Planetary Science, Washington University, 
St Louis, MO}



\begin{abstract}
Stars on the asymptotic giant branch (AGB) produce dust in their 
circumstellar shells.
The nature of the  dust-forming environment is influenced by the evolution of 
the stars, in terms of both chemistry and density, leading to an evolution in 
the nature of the dust that is produced. 
Carbon-rich AGB stars are known to produce 
silicon carbide (SiC). Furthermore, observations of the $\sim$11$\mu$m SiC 
feature 
show that the spectral features change in a sequence that correlates
with stellar evolution.
We present new infrared spectra of amorphous SiC and show that the 
$\sim9\mu$m feature seen in both emission and 
absorption, and correlated with trends in the $\sim$11$\mu$m feature, 
may be due to either amorphous SiC or to nano-crystalline diamond with a
high proportion of Si substituting for C.
Furthermore, we identify SiC absorption in three ISO spectra of extreme 
carbon stars, in addition to the four presented by \citet{speck97}. 
An accurate description of the sequence in the IR spectra of carbon stars
requires accounting for both SiC emission and absorption features. This
level of detail is needed to infer the role of dust in evolution of carbon
stars.
Previous attempts to find a sequence in the infrared spectra of carbon
stars considered SiC emission features, while neglecting SiC absorption
features, leading to an interpretation of the sequence inadequately describes 
the role of dust. 
We show that the evolutionary sequence in carbon star spectra
is consistent with a grain size evolution, such that dust grains get 
progressively smaller as the star evolves. 
The evolution of the grain sizes provides a natural explanation for the shift 
of the $\sim$11$\mu$m SiC feature in emission and in absorption.
Further evidence for this scenario is seen in both post-AGB star spectra and 
in meteoritic studies of presolar grains.

\end{abstract}



%


\section{Introduction}

\subsection{Evolution during the Asymptotic Giant Branch (AGB) phase}

Stars in the mass range 1-8 M$_\odot$ eventually evolve up the asymptotic giant
branch (AGB) where they begin to lose mass and form circumstellar shells of 
dust and gas. During their ascent of the AGB, mass-loss rates are expected 
to increase with the luminosity of the star, 
resulting in progressive optical thickening of circumstellar shells. 
Furthermore,  the chemistry of these stars' atmospheres, and thus of their 
circumstellar shells, changes as a result of dredge-up of newly 
formed carbon from the He-burning shell. Due to the extreme ease of formation 
and stability of CO molecules, the chemistry in circumstellar shells is 
controlled by the C/O ratio. If C/O $<$ 1 all the carbon is trapped in CO, 
leaving oxygen to dominate the chemistry. Conversely, if C/O $>$ 1 all the 
oxygen is trapped in CO, and carbon dominates the chemistry. Stars start their 
lives with cosmic C/O ratios ($\approx$0.4), and are thus oxygen-rich. In some 
AGB stars, the dredge-up of newly formed carbon is efficient enough to raise 
the  C/O above unity, and these stars are known as carbon stars. They are 
expected to have circumstellar shells dominated by amorphous or graphitic 
carbon grains, although other dust grains are also important (e.g.\ 
silicon carbide; SiC). 

As the mass-loss rate increases with the evolution of the AGB stars, the dust 
shells get thicker (both optically and geometrically) 
and these stars eventually become invisible at optical 
wavelengths and very bright in the infrared (IR). Such stars are known as 
``extreme carbon stars''  \citep{volk92,volk00}. 
At this stage, intense mass loss depletes the remaining hydrogen in the 
star's outer envelope, and terminates the AGB. Up to this point the star has 
been making and dredging-up carbon, thus these shells have even more carbon 
available for dust production than optically bright (early) carbon stars 
(i.e.\ the Si/C ratio decreases with the evolution of these stars).
The rapid depletion of material from the outer envelope of the star means that 
this extremely high mass-loss phase must have a relatively short duration 
\citep[a few $\times 10^4$ years;][]{volk00}.

\subsection{Post-AGB evolution}

Once the AGB star has exhausted its outer envelope, the AGB phase ends. 
During this phase the mass loss virtually stops, and the circumstellar shell 
begins 
to drift away from the star. At the same time, the central star begins to 
shrink and heat up from 3000 K until it is hot enough to ionize the 
surrounding gas, at which point the object becomes a planetary nebula (PN). 
The short-lived post-AGB phase, as the star evolves toward to the PN phase, is 
also known as the proto-planetary nebula (PPN) phase. 
The detached dust shell drifts away from the central star 
causing a PPN to have cool infrared colors from its cooling dust shell.
Meanwhile, the optical depth of the dust shell decreases as it expands 
allowing the central star to be seen and making such objects optically bright.
The effect of decreasing optical depth and cooling dust temperatures changes 
the spectrum of the circumstellar envelope, revealing features that were 
hidden during the AGB phase.

\subsection{Observing dust around carbon stars}

Circumstellar shells of carbon stars are expected to be dominated by 
amorphous or graphitic carbon grains. These dust species do not have 
diagnostic infrared features and merely contribute to the dust continuum 
emission. However, silicon carbide (SiC) does exhibit a strong infrared 
feature, which can be exploited when studying carbon stars.

In the context of stardust, SiC has been of great interest since 
its formation was predicted by condensation models for carbon-rich 
circumstellar regions \citep[e.g.][]{fried69,gilman69}. \citet{gilra71,gilra72}
predicted that SiC should re-emit absorbed radiation as a feature in the 
10--13$\mu$m region. Following these predictions, a broad infrared (IR) 
emission feature at $\sim$11.4$\mu$m that is observed in the spectra of many 
carbon stars has been attributed to solid SiC particles
\citep[e.g.][]{hack72,treff74}. Indeed, SiC is now 
believed to be a significant 
constituent of the dust around carbon stars. Although \citet{goebel95} and 
\citet{clem03} adopted the name 11+$\mu$m feature to reflect the 
variation in observed peak positions in astronomical spectra, this implies 
that the the features always occur longward of 11$\mu$m. We have adopted 
$\sim$11$\mu$m to reflect the variations in peak positions of this feature.

The effect of the evolution of the density of the dust shell on observed 
features, and particularly on the $\sim$11$\mu$m feature, have been discussed 
extensively 
\citep[e.g.][see 
\S~\ref{LRS}]{cohen84,baron87,willem88,ck90,goebel95,speck97,sloan98}. 
Not all these previous works agree.
All studies concur that the increasing optical depth leads to decreasing color 
temperatures from the stars as the stellar photosphere becomes hidden from 
view, and the dust from which we receive light becomes progressively cooler. 
At the same time, the $\sim$11$\mu$m feature tends to become weaker (relative 
to the continuum) and flatter topped (less sharp peaked) and possibly broader. 
Various explanations of this behavior have been proposed, but none are 
entirely satisfactory. Initially the diminution of the 
$\sim$11$\mu$m feature was attributed to coating of the grains by amorphous 
carbon 
\citep[e.g.][]{baron87,ck90}. However, meteoritic data does not support
this hypothesis (see \S~\ref{meteor}). Moreover, more recent studies of the 
$\sim$11$\mu$m 
feature have shown that it is consistent with SiC self-absorption, i.e. 
absorption by cooler SiC particles, located in the outer part of the dust 
shell, where they can absorb the SiC emission feature produced by warmer SiC 
closer to the central star (Speck et al. 1997). Indeed, Speck et al. (1997) 
showed that every star in their sample whose underlying continuum temperature 
(T$_{col}$) $<$ 1200~K  was best fitted by self-absorbed SiC, whether the 
$\sim$11$\mu$m feature was in net emission or net absorption. 
As the dust shell reaches extreme optical depths, the $\sim$11$\mu$m feature 
will 
eventually be seen in net absorption. However, these absorption features are 
rare and have mostly been ignored in discussions of 
evolutionary sequences in carbon star spectra.

\subsection{Meteoritic Evidence \label{meteor}}
Silicon carbide is an important presolar grain found in meteorites. 
The isotopic compositions of presolar grains indicate that they originated 
outside the solar system. Presolar SiC was first discovered by \citet{bern87} 
and since then has been the focus of a great deal of laboratory work 
\citep[see][and references therein]{bern05}. 
The most important findings of this work are:
(1) most of the SiC presolar grains were formed around carbon stars; 
(2) nearly all ($\sim$90\%) are of the cubic 
$\beta$-polytype\footnote{Silicon carbide can form into numerous different 
crystal structures, known as polytypes. $\beta$-SiC is the cubic polytype, and 
is now believed to be the dominant form of SiC forming around carbon stars; 
see \S~\ref{meteor}, and \citet{speck97,speck99,daulton03}}; 
(3) with one exception, SiC grains are never found in the cores of carbon 
presolar grains (unlike other carbides - TiC, ZrC, MoC); 
(4) the grain-size distribution is large (1.5nm to 26$\mu$m), 
with most grains in the 0.1 -- 1$\mu$m range, but with single-crystal grains 
can exceed 20$\mu$m in size.

Observations of the $\sim$11$\mu$m feature have been compared with 
laboratory spectra of various forms of SiC and after some false starts, has 
now been attributed to $\beta$-SiC, matching the information retrieved from 
meteoritic samples \citep{speck99,clem03}. However, there are still some 
discrepancies between observational and meteoritic evidence
(most notably related to grain size). 
Studies of meteoritic SiC grains can aid our understanding of their evolution.

\subsection{Purpose}

The goal of this paper is to extract the evolution of the dust from the
infrared spectra of carbon-rich AGB stars (both optically 
bright and ``extreme'') and post-AGB stars. In \S~\ref{lab} 
we present new 
infrared spectra of amorphous silicon carbide which may explain some of the 
features seen in the carbon star spectral sequences.
In \S~\ref{spec} we review studies of the infrared spectra of 
carbon stars which demonstrate the changes in spectral features associated 
with stellar evolution. We also present new detections of $\sim$11$\mu$m 
absorption features in extreme carbon stars, which are essential to our 
understanding of the dust grain evolution. 
In \S~\ref{selfabs} 
we present a
dust evolution sequence to explain the spectral changes in carbon stars.
The summary and conclusions are in \S~\ref{conc}.

\section{New spectra of amorphous silicon carbide \label{lab}}
\subsection{Experimental samples and techniques}

Superior Graphite donated the bulk $\beta$-SiC. The purity is 99.8\%. Bulk 
$\alpha$-SiC (purity = 99.8\%) in the 6H polytype was purchased from Alfa/Aesar
(Lot \# c19h06).  Grain size varied from ~1 to 25 $\mu$m. Nanocrystals of 2-5 
nm size produced by gas-phase combustion \citep{axel96} best matches 
the X-ray diffraction pattern of $\beta$-SiC but contains some 6H as well 
\citep[for details see][]{hofm00}. 
This sample may also contain SiC in the diamond structure, as 
synthesized by \citet{kimura03} and discussed further below. We 
designate this sample as ``nano $\beta+\alpha$''. These three samples were 
studied by \citet{speck99} and \citet{speck04}. Nanocrystals were 
also purchased from Nanostructured and Amorphous Materials, Inc.  
Amorphous SiC of 97.5\% purity consists of 10 $\times$ 100 nm$^2$ laths 
(stock \# 4630js).  Nano-$\beta$ of 97\% purity consists of 20 nm particles 
(stock \# 4640ke). Diamond of 95\% purity consists of 3 nm particles 
(stock \# 1302jgy). Most SiC samples were dark grey, indicative of excess C or 
Si, e.g., inclusions of these elements. Nano-diamond is light grey, presumably 
due to graphite impurities. The $\alpha$-SiC is pale grey to amber and has 
little in the way of impurities.
	
Mid-IR spectra were obtained from $\sim$450 to 4000 cm$^{-1}$ 
(2.2 to 25$\mu$m) at 2 cm$^{-1}$ 
resolution using a liquid-nitrogen cooled HgCdTe detector, a KBr beamsplitter 
and an evacuated Bomem DA 3.02 Fourier transform interferometer. Thin films 
were created through compression in a diamond anvil cell (DAC) which was used 
as a sample holder and for the reference spectrum. Efforts were made to cover 
the entire diamond tip (0.6 mm diameter) with an even layer of sample, but 
slight irregularities in the thickness were inevitable.  Far-IR data were 
obtained from powder dispersed in petroleum jelly on a polyethelene card from  
$\sim$150 to 650 cm$^{-1}$ ($\sim$15 to 67 $\mu$m) using a 
DTGS\footnote{deuterated triglycine sulfate} detector and 
mylar beamsplitter. A 
spectrum of nano-diamond (which should be featureless) was subtracted to 
remove the effect of scattering. For the bulk and nano $\beta + \alpha$ 
sample, a thin film 
was made in a DAC and a helium bolometer served as the detector.  Far- and 
mid-IR spectra were scaled to match in the region of overlap and merged.  
For procedural details see \citet{speck99} or \citet{hofm03}.

\subsection{Laboratory Results}

\begin{figure}
\includegraphics[angle=0,scale=0.8]{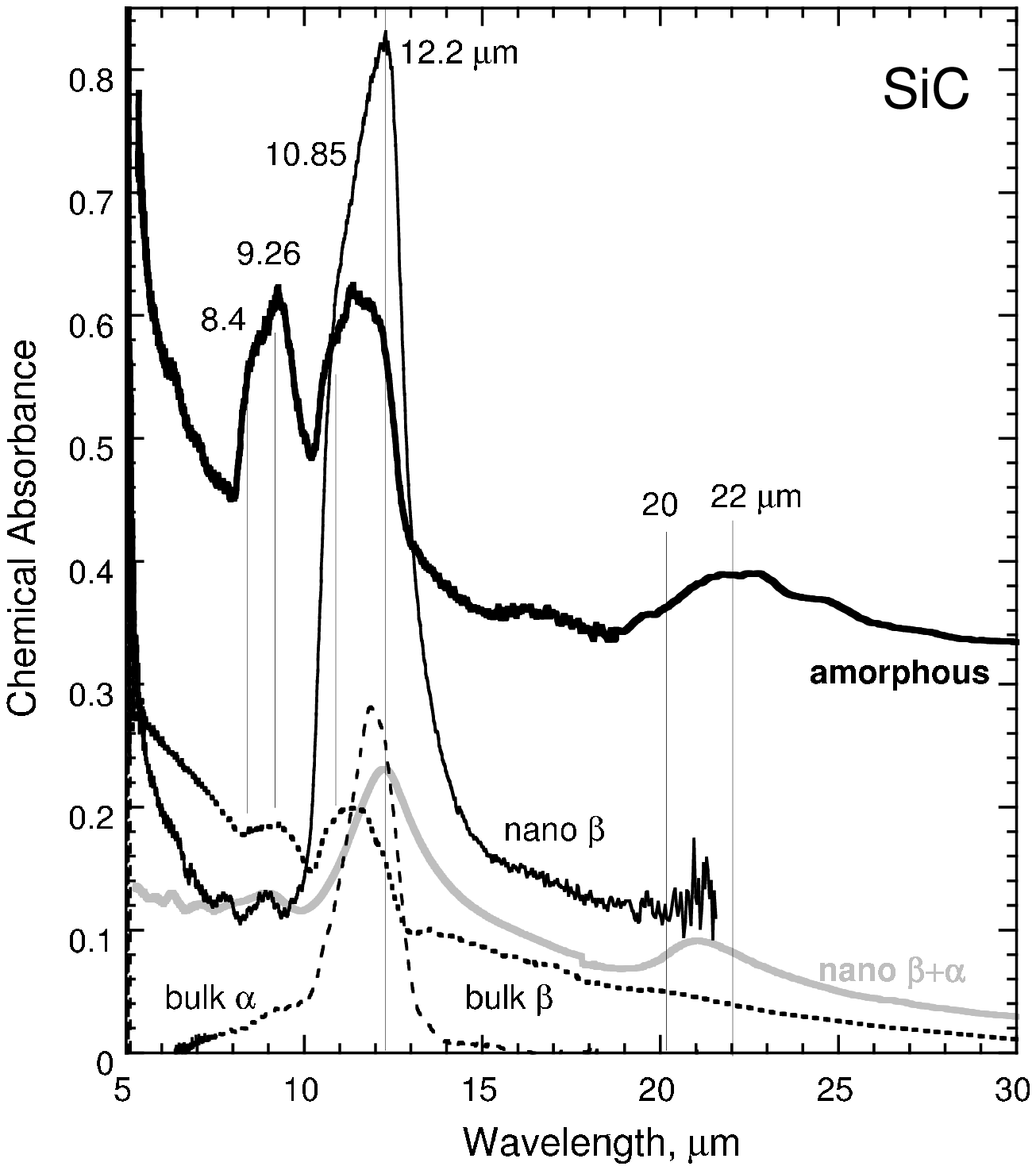}
\caption{Merged far- and mid-IR spectra of various samples of SiC.  
Heavy solid line = amorphous SiC. Light solid line = nano-$\beta$. 
Grey solid line = nano-$\beta + \alpha$. Dotted line = bulk $\beta$.
Dashed line = bulk $\alpha$. Mid-IR segments ($\lambda$= 5 to 25$\mu$m) are 
raw data. Far-IR data was scaled to match, but the merging and scaling are 
equivocal.\label{labdata}}
\end{figure}

The two nanocrystalline samples have spectra that closely resemble each other,
whereas the spectrum of the amorphous sample is most similar to that of bulk 
$\beta$-SiC (Fig.~\ref{labdata}).  
All samples have three peak complexes centered near 
9, 11, and 21$\mu$m. The number and position of the peaks constituting each 
complex vary, and the relative intensities of the complexes vary among the 4 
samples (Table 1).   Bulk $\alpha$-SiC shows only the 11 $\mu$m complex.  
The shoulders in the spectrum shown are due to interference fringes as the 
overtone-combination 
peaks in this area are much weaker \citep[e.g.][]{hofm00}.  
As it is clear from Fig.~\ref{labdata} that the nano 
samples have spectra unlike that of $\alpha$-SiC (and $\alpha$-SiC is absent 
from presolar grains; see \S~\ref{meteor}), the remainder of the 
discussion concerns the amorphous and $\beta$-SiC samples.

\begin{figure}
\includegraphics[angle=0,scale=0.8]{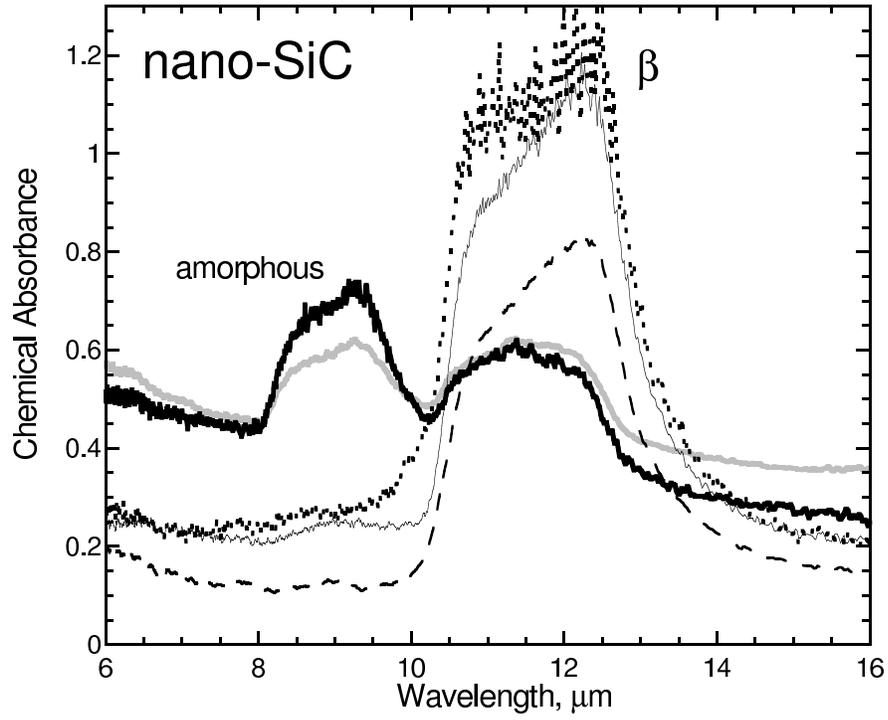}
\caption{Dependence of spectra upon thickness.  
Light lines = nano-$\beta$; 
Heavy lines = amorphous SiC.
Dashed line = thinest nano-$\beta$ sample; dotted line = thickest 
nano-$\beta$ sample; thick black line = thinnest amorphous sample; thick grey 
line = thickest amorphous sample.
Raw data are shown.\label{labthick}}
\end{figure}

Some of the spectral differences are due to grain size. 
From Fig.~\ref{labthick}
\citep[and Fig.\ 2 in ][]{hofm00}, as the sample thickens, the LO 
(longitudinal optic) shoulder increases in intensity relative to the main peak 
(TO, transverse optic).  This occurs because of light leakage 
\citep[see][for a detailed explanation]{hofm03}.  The nanocrystalline samples 
produce 
spectra close to the expected intrinsic profile \citep[see][]{spitz} 
because their very fine grain sizes aid production of a thin, uniform film 
that covers the diamond tip. The rounded profiles for $\beta$- and the 
amorphous SiC are attributed to production of less perfect films from these 
much larger particle sizes. The difference in relative intensity of the 11 and 
9 $\mu$m peaks between the two spectra acquired for amorphous SiC may be due 
to different film thicknesses and amounts of light leakage as well.  However, 
the relative intensity of this pair varied among the spectrum of the 
$\beta$-SiC sample, but was consistent for nano $\beta+\alpha$ 
\citep[Fig. 2 from ][]{hofm00}, 
suggesting an impurity as the origin.

As observed by \citet{speck04} the relative intensity of the 21 and
11 $\mu$m peaks varied among the samples studied.  This observation is 
supported by the two additional samples studied here, and, intensities
of the 21 and 9$\mu$m peaks are roughly correlated (Table 1). 
A feature similar to the 9$\mu$m  peak has been observed in nanocrystals
with varying proportions of Si and C by Kimura and colleagues.  Carbon
films that contain about 30\% Si made by evaporation have a peak at 
9.5$\mu$m \citep{kimura03}.  From electron diffraction, the structure
of the films is nano-diamond and small amounts of $\beta$-SiC are also 
present \citep{kimura03}. Pure diamond structure should not 
have an IR peak, but impurity bands are well known, such as those due to 
nitrogen. Apparently, the Si-C stretch in the diamond structure differs from 
that in the derivative structure of $\beta$-SiC.  
Higher frequency is consistent 
with the lattice constant being smaller for diamond than for SiC.

Ion-sputtered carbon films with proportions of 10, 30, and 50\% Si
similarly contain particles of nano-diamond and  IR absorption bands at 9.5
and 21$\mu$m.  The 30\% film has the most intense 9.5$\mu$m peak, whereas the 
50\% Si film contains nano-crystals of $\beta$-SiC in addition to the
solid-solution nano-diamonds and weak peaks at 11.3 and 12.3$\mu$m
\citep{kimura05a}.

Nano-particles produced by radio-frequency plasma \citep{kimura05b}
absorb at 8.2 (shoulder), 9.2, a doublet at 11.2 and 11.7, and a
moderately intense band at 21$\mu$m.  The structure is nano-diamond.
$\beta$-SiC peaks were rarely seen in the electron diffraction 
results, suggesting that the IR
bands near 11$\mu$m result from short-range order \citep{kimura05b}.
These authors attribute the 21$\mu$m band to excess carbon which may be
present in interstitial sites in the solid-solution nano-diamond particles. 
The present data corroborate and augment their observations.  
Bulk $\beta$-SiC, nanocrystalline material, and 
amorphous nanosamples all behave similarly. 
In the amorphous sample the structural control is lost so that the sample
has places where the $\beta$-SiC structure occurs over nearest neighbors, and
places where the nano-diamond structure occurs. These are about in equal
proportions.  Then there are local sites of excess C giving the 21$\mu$m
band.
That the 21 and 9$\mu$m peaks are strongest in the amorphous material 
corroborates assignment of these peaks to excess C in SiC, one of the 
possibilities proposed by \citet{speck04}.
	
Therefore, the 9$\mu$m peak is indicative of Si-C locally in a diamond 
structure and the 11$\mu$m of Si--C locally in an SiC polytype. 
These peaks are found in amorphous SiC (observed here) and
nano-diamond crystals \citep{kimura03,kimura05a,kimura05b}.
The ratios of 
these peaks are not correlated, but depend on the Si/C ratio of the material. 
Either amorphous SiC or nanocrystalline diamond with a high
proportion of Si substituting for C are good candidates for the carrier of the 
9$\mu$m (see \S~\ref{LRS}) and 21$\mu$m (\S~\ref{21um}) features

\subsection{Comparison with astronomical environments}

\begin{figure}
\includegraphics[angle=0,scale=1.0]{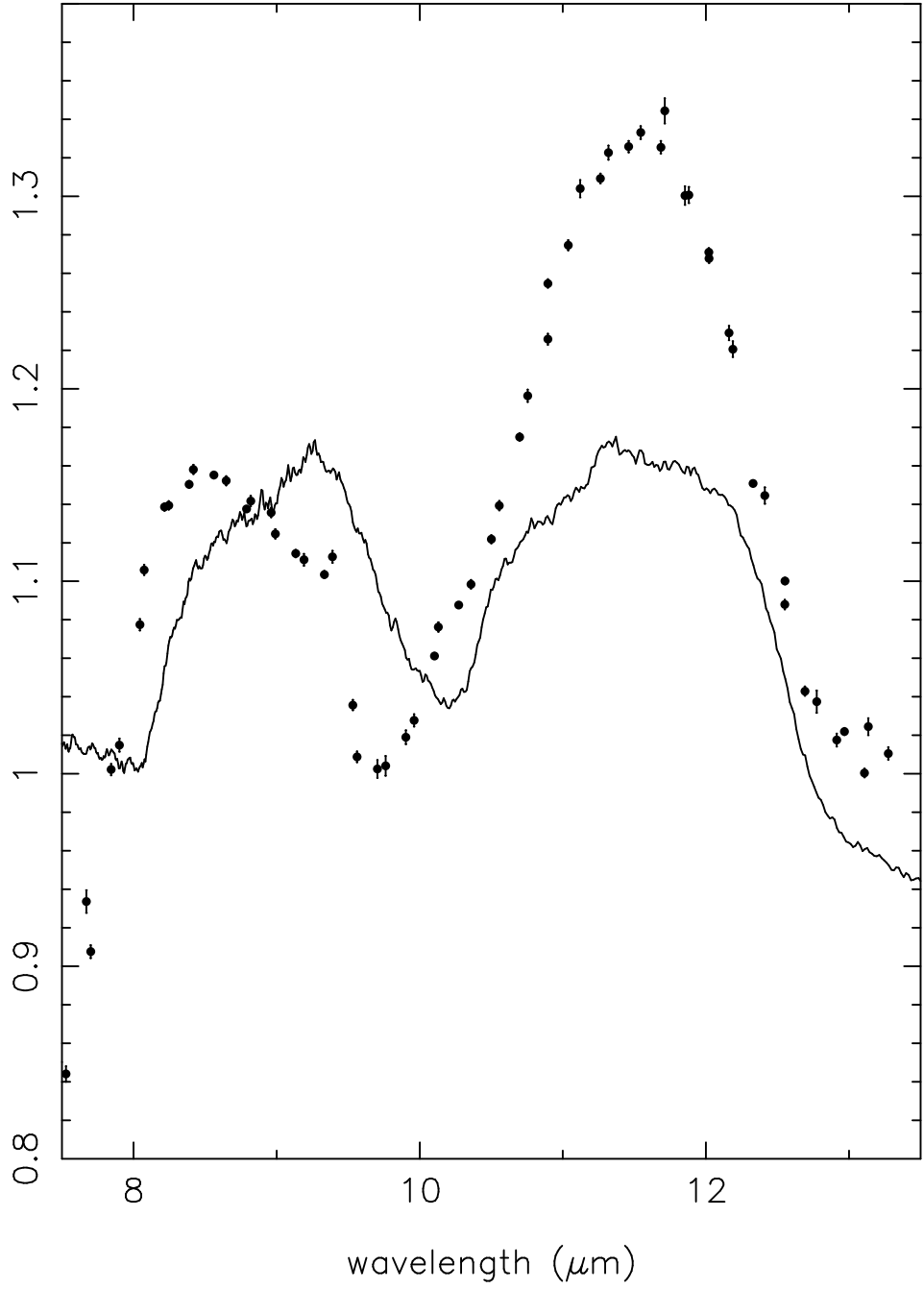}
\caption{Comparison of double-peaked observed astronomical feature in the 
continuum-divided spectrum of Y~CVn (filled circles)  
with laboratory spectrum of amorphous SiC (solid line). \label{ycvn}}
\end{figure}

Fig.~\ref{ycvn} compares the amorphous 
SiC spectrum with the observed,
double-peaked emission feature seen in the spectrum of Y~CVn 
\citep[original published in][]{speck97}.
Amorphous SiC provides peaks at the same positions as those observed in
the  double peaked feature, but not in the same intensity ratio.

\begin{figure}
\includegraphics[angle=0,scale=0.8]{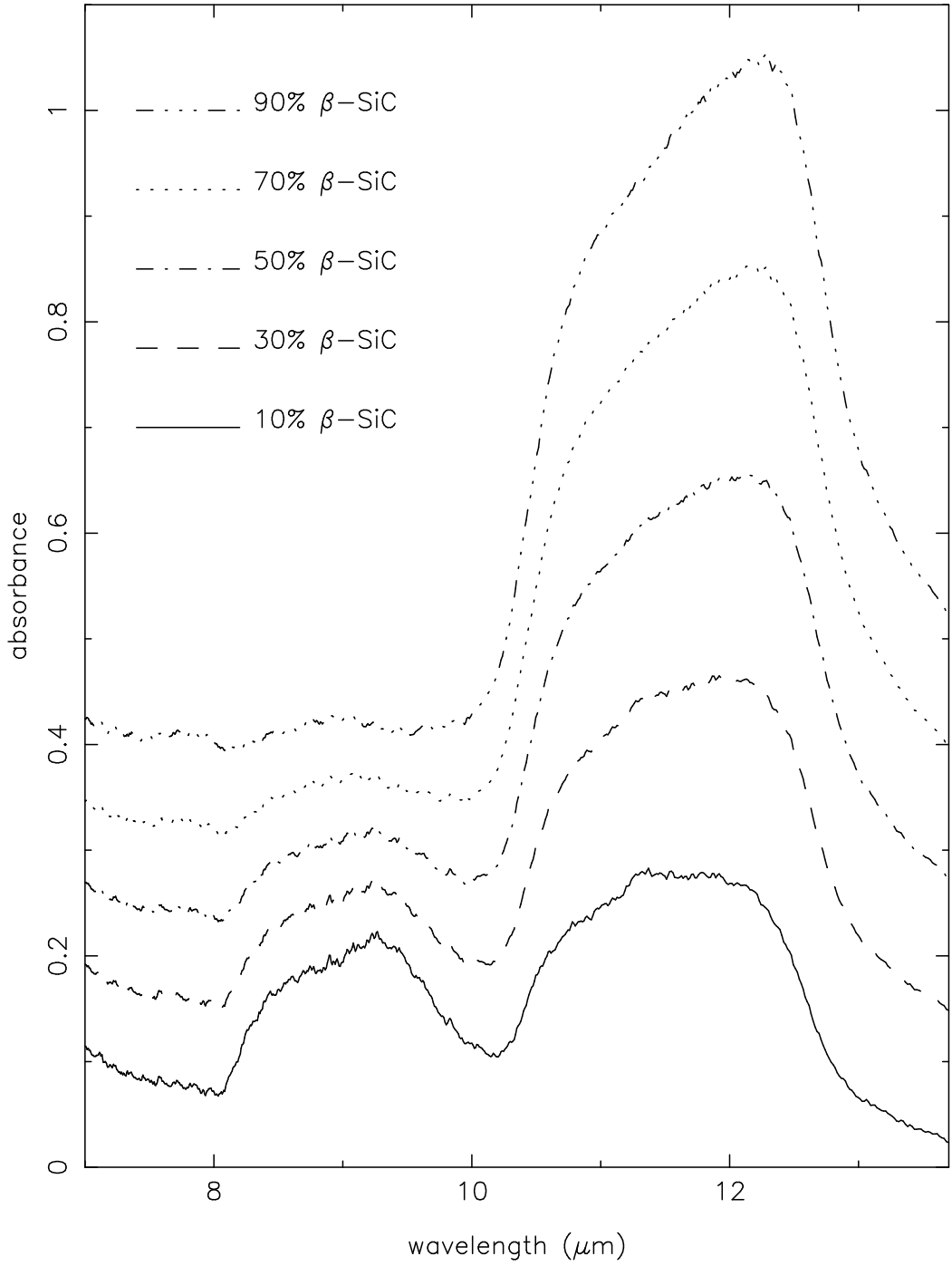}
\caption{The effect of mixing amorphous and $\beta$-SiC. \label{mix}}
\end{figure}

The shorter wavelength peak is due to excess carbon forming diamond-like 
structures in the amorphous or nanocrystalline SiC grains. 
The relative strengths of the two peaks in the 
laboratory spectra are controlled by the concentration of excess carbon in the 
grains. Therefore we can manipulate the composition of the grains to obtain 
bigger or smaller short wavelength peaks, relative to the longer wavelength 
Si--C peak. Furthermore, there may be a contribution to the Si--C part of the 
spectrum from $\beta$-SiC also present in the circumstellar shells, further 
enhancing the longer wavelength peak relative to the shorter one.
 Fig.~\ref{mix} shows the effect of mixing amorphous and crystalline SiC to 
varying degrees.

\begin{figure}
\includegraphics[angle=270,scale=.80]{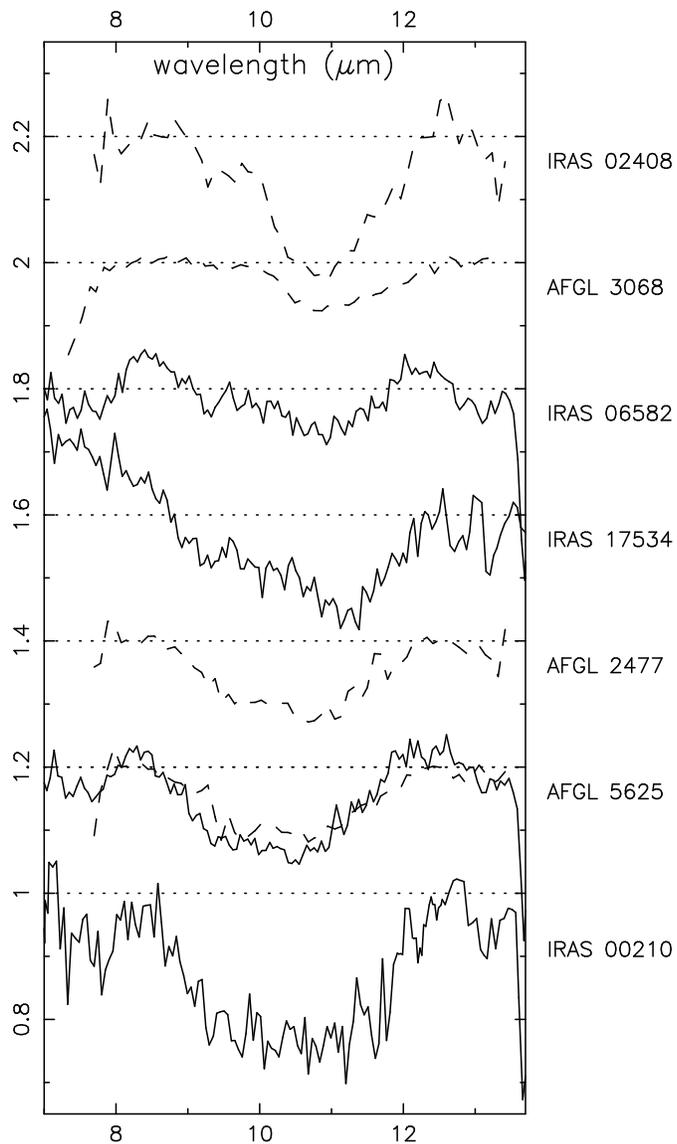}
\caption{Continuum-divided, 7$-$13$\mu$m spectra of extreme carbon stars. 
Solid lines: ISO SWS spectra divided by model fits from Volk et al. (2000). 
Dashed line: UKIRT CGS3 spectra divided by best fit blackbody dust continuum 
from Speck et al. (1997).
Spectra are offset in $y$-direction for clarity: 
IRAS 00210+6221 is not offset; 
AFGL~5625 offset=+0.2;
AFGL 2477 offset=0.4;
IRAS 17534-3030 offset=0.6;
IRAS 06582+1507 offset=0.8; 
AFGL~3068 offset=1.0 
and IRAS 02408+5458 offset=1.2. 
\label{newabsobs}}

\end{figure}

Comparing Figures~\ref{mix} and \ref{newabsobs} shows that the 
double-peaked amorphous SiC spectrum, in combination with $\beta$-SiC can 
explain the variety of absorption features observed in the spectra of extreme 
carbon stars (see also \S~\ref{sicabs}). 
The differences in peak positions between the laboratory spectra and 
the observed astronomical absorption features are due to the effect of 
self-absorption and grain size, which is discussed in \S~\ref{selfabs}.

\section{Spectroscopic studies of carbon stars \label{spec}}

\subsection{{\it IRAS LRS} studies of carbon stars \label{LRS}}

\begin{figure}
\includegraphics[angle=0,scale=0.9]{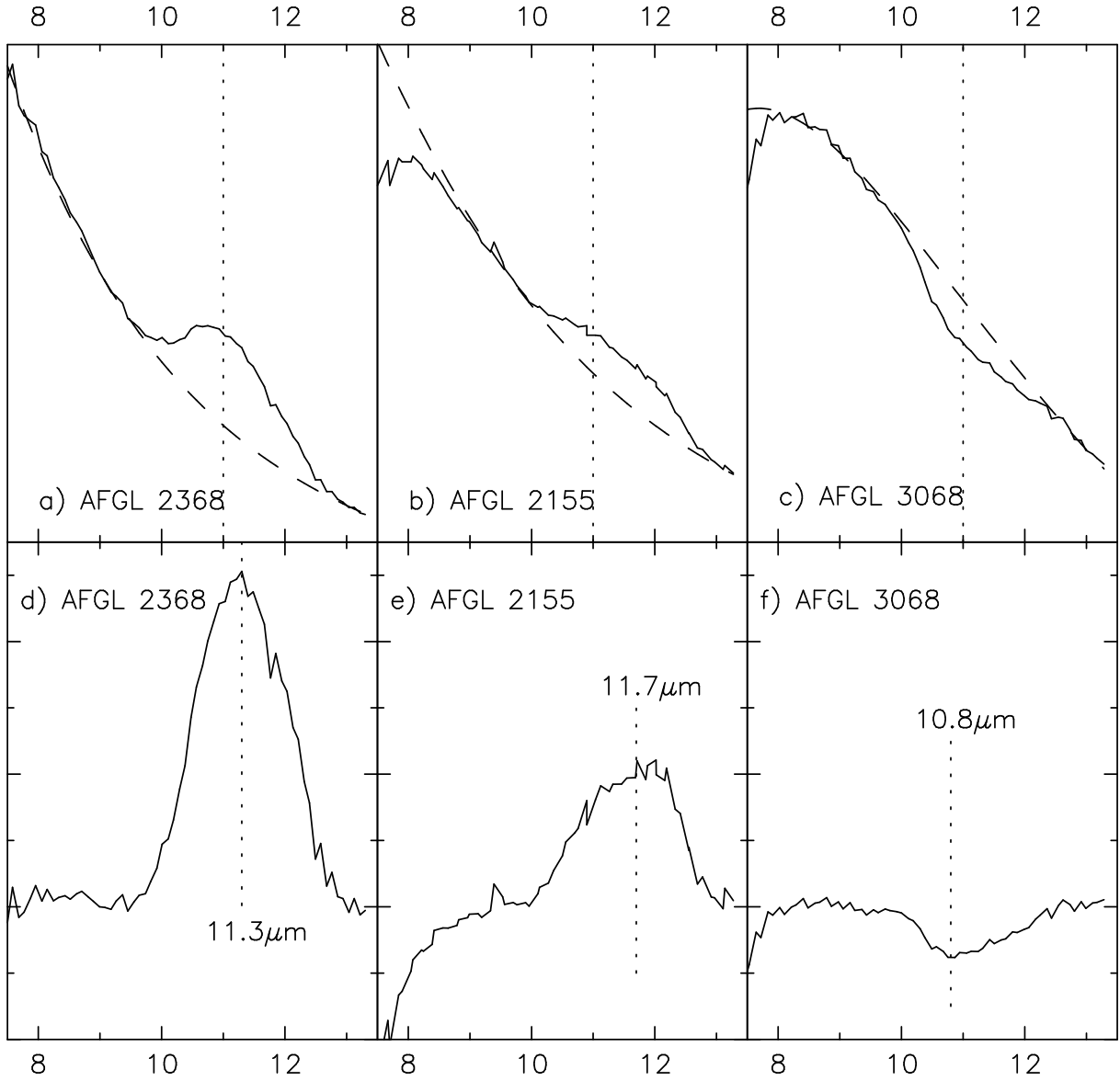}
\caption{UKIRT CGS3 mid-infrared spectra of carbon stars from \citet{speck97}.
{\it Top Row}: Solid line = spectrum (F$_{\lambda}$); dashed line = continuum.
{\it Bottom Row}: continuum divided spectra;
a,d) Pure SiC emission;
b,e) self-absorbed SiC emission;
c,f) SiC (self-)absorption.
$x$-axis is wavelength in $\mu$m. \label{cgs3}}
\end{figure}

The spectra of carbon stars change with the evolution of the star.
Several studies exist of the evolution of the $\sim$11$\mu$m feature in carbon 
star spectra based on {\it IRAS LRS} data, but are somewhat contradictory.
The majority of carbon stars exhibit the $\sim$11$\mu$m feature in emission.
With the exception of 
\citet{speck97}, all attempts to understand the sequence of spectral 
features fail to include the SiC absorption feature. 
Here is the summary of the accepted evolutionary trends:
Early in the carbon star phase, when the mass-loss rate is low and the shell 
is optically thin, the $\sim$11$\mu$m SiC emission feature is strong, narrow 
and sharp. As 
the mass loss increases and the shell becomes optically thicker, the SiC 
emission feature broadens, flattens and weakens. 
Finally, once the mass-loss rate is 
extremely high and the shell is extremely optically thick, the SiC feature 
appears in absorption.
Fig.~\ref{cgs3} shows spectra from \citet{speck97} which demonstrate these 
trends. 
Once the AGB phase ends, the dust thins and cools and we begin to see new 
features that are indicative of the dust in the extreme carbon star phase.

\citet{lml86} found that the majority of $\sim$11$\mu$m emission 
features peak at 
11.15$\mu$m, with only 4\% peaking longward of 11.6$\mu$m.
\citet{baron87} found that, as the continuum temperature decreases and the 
peak-continuum strength of the $\sim$11$\mu$m emission feature diminishes, 
the peak position 
tends to move to longer wavelengths (from $\sim$11.3 to $\sim$11.7$\mu$m). 
This trend is demonstrated in Fig.~\ref{cgs3}.
\citet{willem88} and \citet{goebel95} 
found that the stars with high continuum temperatures tended 
to have the $\sim$11$\mu$m emission feature at 11.7$\mu$m, with the 11.3$\mu$m 
feature arising in the spectra of stars with cool continuum temperatures, 
an obvious 
contradiction to the work of \citet{baron87}. 
Speck et al. (1997) used a much smaller sample of ground-based UKIRT
CGS3 spectra, and found no correlation between the peak position of the 
$\sim$11$\mu$m feature and the continuum temperature.

\citet{baron87} and \citet{goebel95} noticed that the decreasing continuum 
temperature is accompanied by emergence of a second spectral feature at 
$\sim9\mu$m.
\citet{sloan98} 
found an anti-correlation between the $\sim9\mu$m feature and the 
decreasing continuum temperature, which is opposite to what was observed by 
both \citet{baron87} and \citet{goebel95}.
With such a diverse set of analyses and interpretations
of what is essentially the same dataset \citep[except for ][]{speck97},
no clear correlation can be discerned between changes in the spectral
features and evolution of the dust.
The differences in interpretation of the {\it IRAS LRS} data may stem 
from differing underestimations of the depth of the molecular 
(HCN and C$_2$H$_2$) absorptions, 
and thus the continuum level \citep{aoki99}.

\subsection{The SiC absorption feature \label{sicabs}}

The prototype for extreme carbon stars is AFGL~3068 which has 
an 
absorption feature at $\sim 11 \mu$m, tentatively attributed to absorption by 
SiC \citep[][]{jones78}. \citet{speck97} reinvestigated 
AFGL~3068 and confirmed the absorption features. In addition, \citet{speck97} 
discovered three more extreme carbon stars (IRAS 02408+5458, AFGL~2477 and 
AFGL~5625) with $\sim11\mu$m absorption features, attributable to SiC. 
Two of these objects (AFGL~2477 and AFGL~5625) exhibited a double peaked 
absorption feature, with the 
$\sim11\mu$m feature accompanied by a shorter wavelength absorption peak at 
$\sim9\mu$m. \citet{speck97} attributed this shorter wavelength peak to 
interstellar silicate absorption along the line of sight, although it is 
possible that this feature is related
to the  $\sim9\mu$m emission feature that accompanies the $\sim$11$\mu$m 
emission 
feature in the spectra of some carbon stars.
In all four cases, the $\sim11\mu$m absorption feature actually ``peaks'' at 
10.8$\mu$m. 

The absorption features of AFGL~3068 and IRAS 02408+5458 were revisited by 
\citet{clem03} who showed that these feature, as seen in ISO SWS spectra, 
were consistent with isolated $\beta$-SiC nanoparticles. 
However they could not fit the 
short wavelength side of the $\sim$11$\mu$m absorption
feature using SiC alone, which may indicate that the $\sim9\mu$m feature 
absorption is intrinsic to these stars and its strength varies. This will be 
discussed further is \S~6.

The double-peaked absorption features of AFGL~2477 and AFGL~5625 were 
revisited by \citet{clem05}. This work presented new infrared spectra of 
silicon nitride (Si$_3$N$_4$) and found a correlation between the observed 
double-peaked feature and the laboratory absorption spectrum. Furthermore, 
they were able to correlate various weaker longer wavelength absorptions in 
the astronomical spectra, with those of Si$_3$N$_4$ observed in the 
laboratory.  However, while the 
match to the longer wavelength features is good, in their laboratory spectra 
the 
relative strength of the $\sim11$ and $\sim9\mu$m features compared to the 
longer wavelength features indicates that Si$_3$N$_4$ is almost 
certainly present but cannot be solely responsible for these $\sim11$ and 
$\sim9\mu$m absorption features.

\citet{volk00} presented the ISO spectra of five extreme carbon stars. They 
produced radiative transfer models, assuming amorphous carbon dust and 
including a way to fit the broad 26--30$\mu$m feature, but without trying to 
fit the absorption features in the 8$-$13$\mu$m range. 
We have divided the ISO spectra 
by their model fits, and the resulting spectrum of the 7$-$13$\mu$m region 
is shown in Fig~\ref{newabsobs}, together with the continuum-divided 
spectra of extreme 
carbon stars from \citet{speck97}. It is clear that IRAS 06582+1507 shows 
the $\sim$11$\mu$m absorption feature seen in AFGL~3068 and IRAS 02408+5458. 
The spectrum of  IRAS 00210+6221 shows a double/broad 
feature similar to that of 
AFGL~5625. IRAS 17534-3030 seems to be intermediate between the two.
Given that the extreme carbon star phase is not expected to last more than a 
few $\times 10^4$ years, it is not surprising that stars which exhibit these 
features are rare. We now have four stars which exhibit the single 
$\sim$11$\mu$m absorption feature, and three which exhibit the broader double 
feature. It is no longer possible to ignore these absorption features 
when trying to understand the evolution of dust around carbon-stars.

The $\sim$9$\mu$m feature/wing appears to correlate with optical 
depth, appearing strongest when the $\sim$11$\mu$m feature is in absorption, 
but also exists when the $\sim$11$\mu$m feature region weakly emits. 
This feature may be due to amorphous SiC 
with excess carbon (see\S~\ref{lab}), in which case, the formation of 
such grains occurs when the dust shell is denser, further up the AGB. 
This makes 
sense, in that the high density shells may form dust grains so fast that the 
atoms do not have time to migrate to the most energetically favored position 
before another atom sticks on. In this way we would expect to form amorphous, 
rather than crystalline grains. If it is due to nanocrystalline grains, this 
further supports the hypothesis that grain sizes derease as mass-loss rates 
increase. Furthermore, early in the life of a carbon star, the 
dust forming regions will have more Si than C for dust formation (the 
majority of the C atoms will be locked into CO molecules). As the star 
evolves, and more carbon is dredged up from deep within the star, there will 
eventually be more C atoms than Si atoms and excess C will get trapped in 
the grains.

The lack of amorphous SiC in meteoritic samples may be due to the relative 
scarcity of this form of SiC compared to the crystalline polytypes. A typical 
AGB star sends dust out into the interstellar medium 
for a few hundred thousand years, but 
the extreme carbon star phase is much shorter lived ($<10^4$ years) and may 
not occur for all carbon stars (absorption features are rare). Therefore we 
would expect there to be much more crystalline (mostly $\beta$-SiC) than 
amorphous SiC grains. Furthermore, there may well be a mixture of amorphous 
and $\beta$-SiC forming in the circumstellar shells of very evolved (extreme) 
AGB stars. Alternatively, the meteoritic data may support the attribution of 
the 9 and 21$\mu$m features to nano-crystalline SiC grains with diamond 
inclusions. Nanometer-sized SiC grains have been found in the presolar SiC 
samples \citep{bern05}.

\subsection{Carbon-rich post-AGB spectra: the 21$\mu$m feature \label{21um}}

Among the C-rich PPNs, approximately half exhibit a feature in their infrared 
spectra at 21$\mu$m \citep{omont95}. Subsequent higher resolution data revised 
the so-called 21$\mu$m position to 20.1$\mu$m \citep{volk99}. PPNs that 
display this feature are all C-rich and all show evidence of {\it s-process} 
enhancements in 
their photospheres, indicative of efficient dredge-up during the ascent of 
AGB \citep{vwr00}. The 21$\mu$m feature is rarely seen in the spectra of 
either the PPN precursors, AGB stars, or in their successors, PNs 
\citep[however, see][]{volk00,hony}. 
The observed peak 
positions and profile shapes of the 21$\mu$m feature are remarkably constant 
\citep{volk99}. This enigmatic feature has been widely discussed since its 
discovery \citep{kwok89} and has been attributed to a variety of both 
transient molecular and long-lived solid-state species, 
but most of these species have since been discarded as carriers, 
except for HACs/PAHs 
\citep{just96,volk99,buss90,grish01} and 
SiC \citep{speck04}. 
In the case of 
SiC, it is necessary for the dust grains to be small and contaminated with 
carbon impurities in order for this feature to appear 
\citep{kimura05a,kimura05b}. 
Furthermore, the 
cooling and thinning of the dust shell is essential to the emergence of this 
feature. During the AGB phase, the dust is too warm for this feature to 
appear, but as the dust cools the $\sim$11$\mu$m feature of SiC is diminished 
and 
the 21$\mu$m feature is promoted by the underlying dust-continuum emission. 
In this case the changing spectrum reflects the change in temperature and 
optical depth of the dust shell, whereas the spectral features are indicative 
of the last stage of carbon-rich evolution (the extreme carbon star 
phase).

\section{Self Absorption: the effect of changing grain sizes \label{selfabs}}

\citet{cohen84} interpreted the change in the appearance of $\sim$11$\mu$m 
feature 
from sharp and narrow  to broad and flat-topped as possibly an effect of 
self-absorption. This was supported by 
\citet{speck97}, who showed that all carbon star spectra in their sample with 
dust continuum temperatures less than 1200~K needed to have self-absorbed SiC 
in order to be fitted well. Self-absorption exhibits some interesting 
characteristics that can be used to diagnose the nature of the dust grains 
that produce observed spectra.

\begin{figure}
\includegraphics[angle=0,scale=1.0]{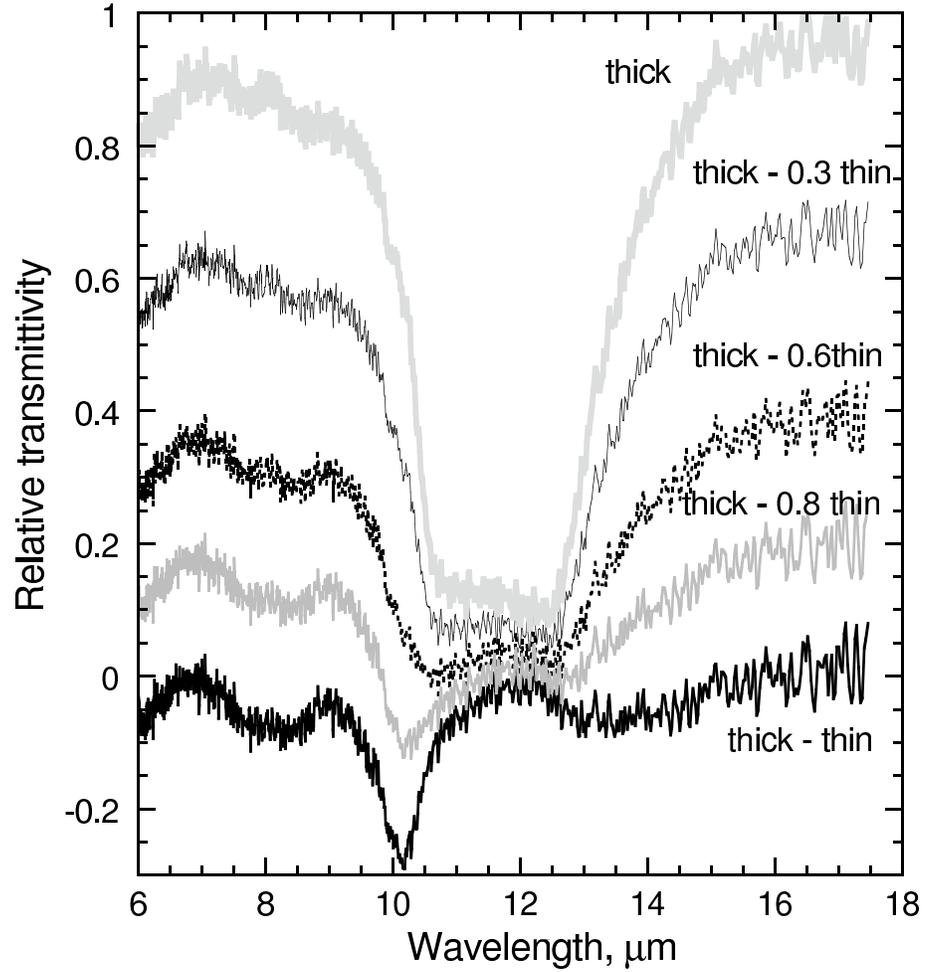}
\caption{The effect of thickening dust samples. The fractional factor is $f$ 
(see Eq.~\ref{eq1}), 
which accounts for more or less light coming from the emitting grains. 
If the grains are all the same size, the absorption feature is in the same 
position as the emission feature. If the absorbing grains are larger, then 
there is an apparent shift in the feature position.
\label{modelthick}}
\end{figure}

Figure~\ref{modelthick} shows how the absorption profile changes as particle 
thickness (i.e.\ grain size) increases.  This appears to be a violation of 
Beer's Law, in that the absorbance is not simply increasing 
(for a single wavelength) with thickness of the particles.
This departure occurs because the
measured intensity depends not only on the amount of light that the sample
absorbs, but also on the amount reflected at the surface facing the
source.  The main peak (the TO mode) becomes saturated in measured spectra
when the actual amount of light transmitted at the TO frequency equals the
reflectivity. This point is reached at lower thicknesses (grain sizes) for the 
strong TO mode than for the more weakly
absorbing shoulder.  Violations of Beer's law will alter self-absorption in
astronomical environments but not emission spectra because the latter are not
affected by surface reflections. Therefore, emission spectra of the grains 
are not so sensitive to the grain size. As long as the grains are still small 
compared to the wavelength at which they are observed 
(i.e. grains smaller than $\approx 1\mu$m for 10$\mu$m spectra),
the pure emission feature will appear at the same wavelength ($\sim11.3\mu$m).
However, once the grains are cool enough and the optical depth high enough, 
self-absorption of the 11.3$\mu$m feature will begin and the profile of the 
feature will depend on the relative sizes of the grains.

In computing the effect of Beer's law violations on
self-absorption, we assume that the following holds: 
(1) the absorbing particles are thicker (larger) than the emitting particles, 
(2) the light received by the absorbing particles is that of the star = 
I$_0$, i.e., the inner cloud of emitting particles is
rarified enough that it contributes negligible intensity compared to the
star, and 
(3) that outer dust transmits more light that the inner dust emits 
(i.e. starlight is also transmitted).

Beer's law states that the wavelength-dependent absorbance $a(\lambda)$ of 
a given substance is proportional 
to the mass absorption coefficient $\kappa_{abs}(\lambda)$, 
and the thickness through which the light has to pass $d$:

\[ a(\lambda) = \kappa_{abs}(\lambda) d \]

Deviations from Beer's law may occur for many reasons. Strictly, Beer's law 
applies to transparent particles 
(i.e.\ particles which transmit some light at all wavelengths)  
and therefore, at high concentrations/opacities may no longer be applicable. 
Other problems include stray light \citep{machof}.

When measuring absorption, we usually measure the transmittivity $T(\lambda)$ 
of a material and thus obtain the absorptivity $A(\lambda)$:

\[ T = \frac{I_{trans}}{I_{0}} = 1 - \frac{I_{abs}}{I_{0}} = 1- e^{-a}\]
\[ A = \frac{I_{abs}}{I_{0}} = e^{-a} \]

The light received from a circumstellar shell $I_{dust}$ is given by:

\[ I_{dust} = I_{emit} + I_{trans} \]

where $I_{emit}$ is the emission from the inner dust and $I_{trans}$ is the 
transmission from the outer dust.

From Kirchhoff's law, the emissivity is the same as the 
absorptivity, so that $I_{emit} = I_{abs}$

\noindent
Therefore,

\[ \frac{I_{dust}}{I_0} = f \frac{I_{abs}}{I_0} + \frac{I_{trans}}{I_0} \]
\[ \frac{I_{dust}}{I_0} = f e^{-a} +  1- e^{-a} \]

\noindent
where the factor $f$ is varied  from 0 to 1 as the fraction of light
originating from emitting dust increases.

The absorption coefficient 
will change with grain size $d$, because of the effect of reflections. 

\begin{equation}
\frac{I_{dust}}{I_0} = f e^{-\kappa_{abs,inner}d_{inner}} +  
1 - e^{-\kappa_{abs,outer}d_{outer}} 
\label{eq1}
\end{equation} 

Essentially, Eq.~\ref{eq1} provides the effective transmission of the dust.  
If Beer's law is followed (i.e. reflectivity is low and the particles are thin
enough to transmit light at all frequencies), 
then $\kappa_{abs,inner} = \kappa_{abs,outer}$ and
the spectrum received is no different than the intrinsic absorptions.  
However, if the absorbing particles are larger, and opaque at some frequencies 
(i.e.\ the TO mode), the spectrum will be altered. We have computed the effect 
of particle thickness on self-absorption in circumstellar shells, and the 
resulting spectra are shown in
 Fig.~\ref{modelthick}. 
For $\kappa_{abs,outer}d_{outer}$, 
we used baseline-corrected
absorption spectra from the thickest sample of nano $\beta$-SiC shown in 
Fig.~\ref{labthick}, and for $\kappa_{abs,inner}d_{inner}$, 
we similarly used the intermediate sample. The same
results would be obtained for the thinnest sample. As the contribution of
the emitting particles increases, the contribution of the LO component
increases relative to the TO component, and the TO component appears to
shift to longer wavelengths.

If the size and absorption coefficients are identical for both the inner and 
outer regions of the dust shell, then there will be no shift in the spectral 
features between emission and absorption. However, if the outer grains are 
larger, there will be a shift in the absorption to shorter wavelengths, 
shown in Fig.~\ref{modelthick}.

\begin{figure}
\includegraphics[angle=270,scale=1.0]{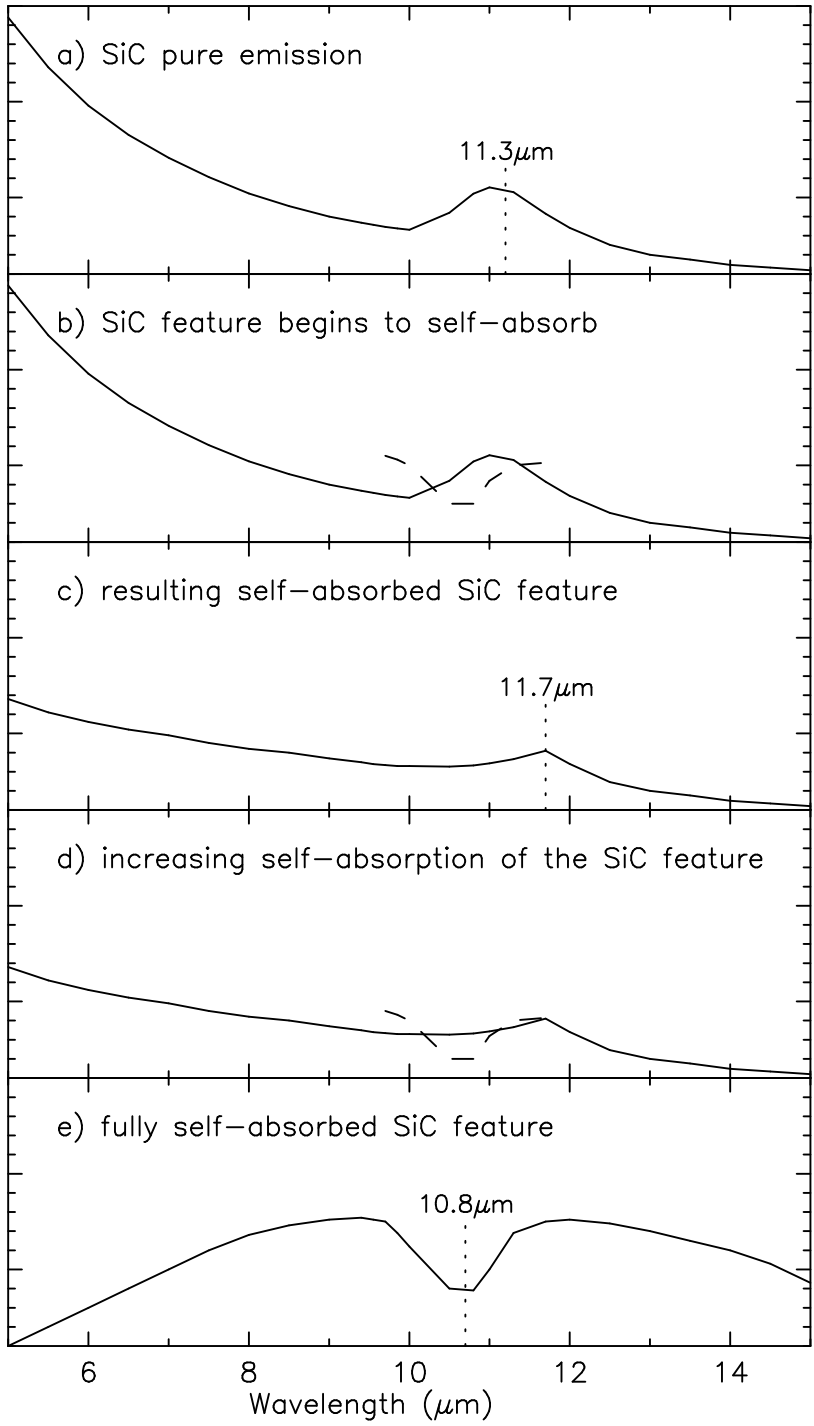}
\caption{Schematic of self-absorption and its effect on the peak postion of 
the $\sim$11$\mu$m feature.
\label{cartoon}}
\end{figure}

In terms of what we would expect to see in the sequence of carbon star 
spectra, the discussion above means that for optically thin dust shells, 
where we are seeing pure emission, the SiC feature should peak at 11.3$\mu$m 
and be sharp. As the optical depth goes up, the SiC feature will become 
self-absorbed, 
but if the grains in the outer part of the shell are larger than 
those in the inner zone, the absorption will occur preferentially on the LO 
side of the feature, diminishing the blue side, which would appear as a shift 
in the feature to longer wavelengths ($\sim$11.7$\mu$m). 
As the optical depth gets 
high enough for the emission feature to be completely absorbed, we no longer 
see the 11.7$\mu$m feature and the absorption will
peak at a shorter wavelength (10.8$\mu$m) than the regular SiC feature at 
11.3$\mu$m. This is shown schematically in Fig.~\ref{cartoon} 
(c.f. Fig.~\ref{cgs3}).
This mechanism explains the evolution 
in the observed SiC absorption feature in extreme carbon stars. 
Furthemore, it correlates with the evolution of the spectral features as seen 
by \citet{baron87}, who showed that as the continuum temperature decreases and 
the peak-continuum strength of the SiC feature diminishes, the peak position 
tends to move from $\sim$11.3$\mu$m to $\sim$11.7$\mu$m. This suggests that 
there is an evolution in the dust grains towards smaller dust 
grains with higher mass loss.

Two separate studies of presolar SiC suggest that there is an evolution in
grain size that corresponds to the evolution of carbon stars. \citet{prombo}
found a correlation between grain size and the concentration of 
{\it s-process} elements in SiC grains taken from the Murchison meteorites. 
The Indarch meteorite presolar SiC grains yielded similar results 
\citep{jennings}. In both cases, the smaller grains have higher relative 
abundances of {\it s-process} elements. Since newly-formed
 {\it s-process} elements are dredged-up along with carbon from the He-burning 
shell, they are more abundant in more evolved circumstellar shells than 
early ones. 
Therefore, these results suggest that
the grains formed around carbon stars decrease in size as the stars evolve and 
support the observations of the self-absorption effects described above. 

Up until now, the generally accepted wisdom has been that 
low mass-loss rates early in the AGB phase lead to small grains, and 
increasing mass-loss rates permit the growth of larger grains. In the new 
scenario, smaller grain sizes with increasing mass loss can be understood in 
terms of potential nucleation sites. The hardest step in grain formation is 
the production of seeds onto which minerals can grow. In a low-density gas 
very few seeds can nucleate, leading to the formation of few grains. However, 
these few grains can grow large because there are not so many grains competing 
for the same atoms. Conversely, in a high-density gas, it is 
easier to form seeds, but so many more nucleate, that there are not enough 
atoms around for any individual grain to grow large. In this way, high 
densities lead to very dense dust shells of small grains. 
This effect may be exacerbated by the earlier outflows moving 
slower than the later outflows.
If the stellar winds that drive the mass loss 
and send the newly formed grains away from star accelerate with time, then the 
earlier grains will have more time closer to the star in regions dense enough 
for grain growth to occur. Since radiation-pressure coupling should be related 
to grain composition, and this will vary with the Si/C ratio, we may expect 
such an effect.

\section{Conclusions \label{conc}}

The dust shells around carbon stars and their successors, carbon-rich post-AGB 
stars, evolve along with the star.
In early phases,  the mass-loss rate is low, and hence the dust 
shell is optically thin. As these stars evolve, the mass-loss rate increases, 
and thus the dust shell gets more optically thick, until the star is 
completely obscured in visible light and very bright in the infrared. 
During this evolution the relative abundance of silicon to carbon (Si/C) 
available for dust formation decreases, so that in the early carbon stars 
there is more Si and C, whereas in the extreme carbon stars there is much
more C than Si. Once the AGB phase ends, the existing dust shell spreads out, 
becoming optically thin and cooler.
The combination of increasing density and increasing carbon on the AGB 
manifests itself 
in the nature of the dust grains as seen in the spectral sequence for carbon 
stars.

Early in the carbon star phase, when the mass-loss rate is low and the shell 
is optically thin, the $\sim$11$\mu$m SiC emission 
feature is strong, narrow and sharp. As 
the mass loss increases and the shell becomes optically thicker, the SiC 
feature broadens, flattens and weakens. Finally, once the mass-loss rate is 
extremely high and the shell is extremely optically thick, the SiC feature 
appears in absorption.
The shifts in the peak of the $\sim$11$\mu$m SiC feature are attributable to a 
combination of optical depth and grain sizes.
The only way to see a shift in the absorption feature to 10.8$\mu$m is if 
the absorbing grains are larger than the emitting grains.
Therefore, we have observational evidence to suggest that the grains formed 
in circumstellar shell get smaller as the star evolves. Further evidence for 
this scenario is seen both in post-AGB spectra and in meteoritic studies 
of presolar grains.

We have presented new mid-IR laboratory spectra for various forms of SiC 
including amorphous SiC with excess carbon structure. 
These data corroborate and augment the laboratory work of 
\citet{kimura03} and \citet{kimura05a,kimura05b}, 
indicating that solid solutions of C 
replacing Si in SiC have the diamond 
structure on a local scale.
This is a good candidate
for the carrier of the $\sim9\mu$m feature seen in both emission and 
absorption, and correlated with trends in the $\sim$11$\mu$m feature.

\acknowledgments
We are extremely grateful to the reviewer whose comments significantly 
improved the paper. We would also like to thank Kevin Volk for providing us 
with his ISO SWS spectra and model fits for extreme carbon stars.
Support for AMH was provided by NASA grant APRA04-0000-0041.




\clearpage



\clearpage
\clearpage

\clearpage

\clearpage

\clearpage

\clearpage

\clearpage

\clearpage

\begin{table}
\begin{center}
\caption{Spectral parameters$^\ast$ of SiC in various forms. \label{tbl-1}}
\begin{tabular}{lccccc}
\tableline\tableline
                      & bulk $\beta$ & amorphous  & nano-$\beta$ & nano-$\beta + \alpha$ & bulk $\alpha$\\
\tableline
Grain size (nm)       & $>$1000        & 10 -- 100  & 3            & 2 -- 5                & ---\\
shoulder (LO; $\mu$m) & 8.4          & 8.4        & ---          & ---                   & 7.66 (sharp)\\
main peak ($\mu$m)    & 9.3          & 9.26       & 8.95         & 9.0                   & ---\\
shoulder (LO; $\mu$m) & 10.58        & 10.6       & 10.85        & ---                   & 10.55\\
Barycenter or main ($\mu$m)& 11.30   & 11.43      & ---          & 12.14                 & 11.85\\
Main peak or shoulder($\mu$m)& 12.1  & 12.2       & 12.27        & 12.14                 & 12.23\\
peak ($\mu$m)         & ---          & 16.3       & ---          & ---                   & 16.02 (sharp)\\
shoulder ($\mu$m)     & ---          & 20         & ---          & ---                   & ---\\
main peak ($\mu$m)    & 20.5         & 22         & ---          & 21                    & ---\\
I$_{9}$/I$_{11}$      & $\sim$0.35   & 0.73       & 0.050        & 0.15                  & ---\\
I$_{21}$/I$_{11}$     & $\sim$0.1    & $\sim$0.26 & ---          & $\sim$0.3$^{\dag}$    & ---\\
\tableline
\end{tabular}
\begin{tabular}{p{6.5in}}
$\ast$ Peak positions obtained from thinnest samples examined.\\
$\dag$ Intensity is very uncertain as the $\sim  21 \mu$m peak was measured using a dispersion\\
\end{tabular}
\end{center}
\end{table}



\clearpage

\end{document}